\documentclass[%
reprint, nature
]{revtex4-2}

\usepackage{graphicx}
\usepackage{amsmath,amssymb}
\usepackage{xfrac}
\usepackage{siunitx}
\usepackage[english]{babel}
\usepackage{relsize}
\usepackage{float}
\usepackage{hyperref}
\usepackage{braket}

\DeclareMathOperator{\sech}{sech}

\begin{document}
	
	\title{{Time-bin entangled Bell state generation and tomography on thin-film lithium niobate}}
	
	\author{Giovanni Finco}\email{gfinco@phys.ethz.ch}
	
	\author{Filippo Miserocchi}
	\author{Andreas Maeder}
	\author{Jost Kellner}
	\author{Alessandra Sabatti}
	\author{Robert J. Chapman}
	\author{Rachel Grange}
	
	\affiliation{ETH Zurich, Department of Physics, Institute for Quantum Electronics, Optical Nanomaterial Group, Auguste-Piccard-Hof, 1, 8093, Zurich, Switzerland}
	
	\date{\today}
	
	\begin{abstract}
		Optical quantum communication technologies are making the prospect of unconditionally secure and efficient information transfer a reality. The possibility of generating and reliably detecting quantum states of light, with the further need of increasing the private data-rate is where most research efforts are focusing. The physical concept of entanglement is a solution guaranteeing the highest degree of security in device-independent schemes, yet its implementation and preservation over long communication links is hard to achieve. Lithium niobate-on-insulator has emerged as a revolutionising platform for high-speed classical telecommunication and is equally suited for quantum information applications owing to the large second-order nonlinearities that can efficiently produce entangled photon pairs. In this work, we generate maximally entangled quantum states in the time-bin basis using lithium niobate-on-insulator photonics at the fibre optics telecommunication wavelength, and reconstruct the density matrix by quantum tomography on a single photonic integrated circuit. We use on-chip periodically-poled lithium niobate as source of entangled qubits with a brightness of \SI{242}{\MHz/\mW} and perform quantum tomography with a fidelity of \SI{91.9}{}$\pm$\SI{1.0}{\%}. Our results, combined with the established large electro-optic bandwidth of lithium niobate, showcase the platform as perfect candidate to realise fibre-coupled, high-speed time-bin quantum communication modules that exploit entanglement to achieve information security.
	\end{abstract}
	
	\maketitle
	
	\section*{Introduction}
	Quantum information technologies are in need of stable and reliable modules for generation and detection of entangled states \cite{orieuxRecentAdvancesIntegrated2016}. Practical quantum communication schemes make use of the already deployed optical fibre links and have been proved secure to transmit quantum information over more than \SI{500}{\km} \cite{chenTwinfieldQuantumKey2021}. Traditionally, polarisation states of light were exploited, yet random fibre birefringence along communication links renders polarisation encoding impractical to implement on standard telecom networks, therefore spatial mode multiplexing  \cite{dingHighdimensionalQuantumKey2017} or time-bin encoding are preferred due to their robustness against polarisation scrambling \cite{brendelPulsedEnergyTimeEntangled1999}. Between these two options, time-bin encoding further simplifies the scheme as it can use standard optical fibres instead of multi-core links or sophisticated deconvolution algorithms to distinguish qubits. To streamline the complex task of generating, manipulating and detecting quantum states while maintaining coherence over long communication links, weak laser pulses are typically adopted instead of single photon sources, and decoy states are implemented to enhance security \cite{gisinQuantumCommunication2007, sasakiFieldTestQuantum2011, dynesCambridgeQuantumNetwork2019, paraisoPhotonicIntegratedQuantum2021, dolphinHybridIntegratedQuantum2023} following the BB84 protocol \cite{bennettQuantumCryptographyPublic2014}. Among the several proposed approaches for quantum key distribution (QKD), the E91  \cite{ekertQuantumCryptographyBased1991} and BBM92 \cite{bennettQuantumCryptographyBell1992} protocols rely on entanglement to achieve unconditional security. In an entangled QKD scheme, quantum correlations are used to ensure security of the channel and detect eavesdroppers, as a tentative hack of the communication leads to introduction of classical correlations which can be detected by assessing that entanglement has been disrupted instead of implementing more cumbersome bit-error checks as in BB84 \cite{bennettQuantumCryptographyPublic2014, ekertQuantumCryptographyBased1991}.
	
	As for classical telecommunication, it naturally follows that quantum communication systems also require integrated solutions based on optoelectronic components; these are necessary when aiming at large-scale production and global quantum network deployment \cite{ramakrishnanIntegratedPhotonicPlatforms2023, luoRecentProgressQuantum2023}. Time-bin encoding and Franson interferometry \cite{fransonBellInequalityPosition1989} have been implemented on photonic integrated circuits as part of the efforts to miniaturise communication elements towards fibre-compatible quantum information technologies \cite{xiongCompactReconfigurableSilicon2015, zhangIntegratedSiliconNitride2018, thielTimebinEntanglementTelecom2024}. Among the available platforms able to accomplish the task, the most widespread are silicon-based technologies \cite{dingHighdimensionalQuantumKey2017,  sibsonIntegratedSiliconPhotonics2017, zhangIntegratedSiliconPhotonic2019, avesaniFullDaylightQuantumkeydistribution2021}, silica \cite{politiSilicaonSiliconWaveguideQuantum2008, saxHighspeedIntegratedQKD2023}, 
	or III-V semiconductors; the latter being used as a source of entangled photons \cite{chenInvitedArticleTimebin2018} or as pump laser diodes in a hybrid integration scheme \cite{sibsonChipbasedQuantumKey2017, shams-ansariElectricallyPumpedLaser2022, dolphinHybridIntegratedQuantum2023}. Vacancy centres or defects in diamond and silicon as well as quantum dots are also being successfully used as sources for applications in quantum information \cite{suHighperformanceDiamondbasedSinglephoton2009, jayakumarTimebinEntangledPhotons2014, lodahlInterfacingSinglePhotons2015, borregaardOneWayQuantumRepeater2020, bathenManipulatingSinglePhotonEmission2021, knallEfficientSourceShaped2022}. 
	
	\begin{figure*}[t!]
		\centering
		\includegraphics[width=\textwidth]{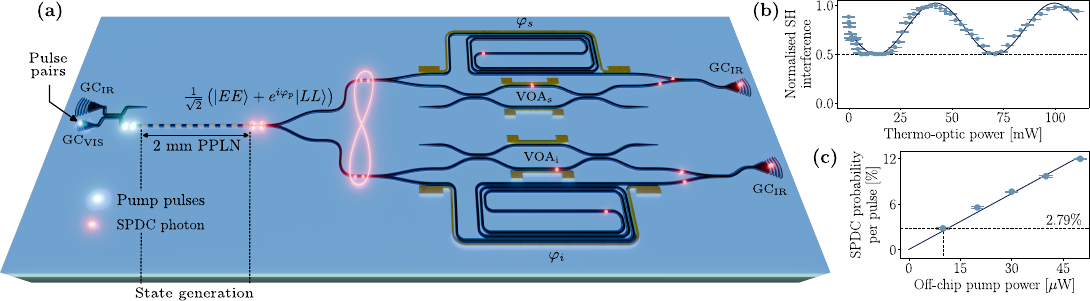}
		\caption{\textbf{Device, working principle and characterisation.} (a) Pump pulses are generated off-chip and input from the left via grating couplers; they undergo SPDC to generate the desired Bell state and probabilistically split to two twin Franson interferometers for tomography. Entanglement is maintained when photons propagate across two analysis interferometers where projection onto the desired bases occurs. $GC$: grating coupler, $VOA$: variable optical attenuator, $\varphi$: projection phase, $s$: signal, $i$: idler. (b) Normalised second-harmonic interference measured by propagating a \SI{1550}{\nm} femtosecond laser backwards through device and setup. (c) SPDC generation probability versus off-chip average pump power.}
		\label{fig:01_Device}
	\end{figure*}
	
	Lithium niobate-on-insulator (LNOI) is rapidly emerging as a leading alternative for classical communication applications thanks to its low propagation loss, wide transparency range and large electro-optic (EO) bandwidth \cite{zhuIntegratedPhotonicsThinfilm2021}. Fabrication of high-quality photonic circuits on LNOI is now established and reliable, with low-loss optical circuits being routinely produced for various applications \cite{ shams-ansariReducedMaterialLoss2022, kaufmannRedepositionfreeInductivelycoupledPlasma2023}. EO modulators have been demonstrated on LNOI to reach several tens of \SI{}{\GHz} and are now commercially available \cite{zhangIntegratedLithiumNiobate2021, pohl100GBdWaveguideBragg2021}. Ultra-broadband EO-combs \cite{wangMonolithicLithiumNiobate2019, zhangPowerefficientIntegratedLithium2023}, generation of short pulses \cite{yuIntegratedFemtosecondPulse2022} as well as phase shifters and switches \cite{maederHighbandwidthThermoopticPhase2022} have also been demonstrated. Additionally, the non-centrosymmetric structure of single crystal lithium niobate films allows for efficient second-order frequency conversion processes by domain engineering and periodic poling \cite{wangUltrahighefficiencyWavelengthConversion2018}. These aspects make LNOI an ideal platform for the next generation of optical communication devices beyond the classical domain, as spontaneous parametric down-conversion (SPDC) can be leveraged to generate entangled photons at the different wavelengths with exceptionally high brightness \cite{saraviLithiumNiobateInsulator2021, zheng16LithiumNiobate2023, harperHighlyEfficientVisible2024}. SPDC is advantageous over spontaneous four-wave mixing (SFWM) not only because of its higher efficiency, but also because the down-converted photons are spectrally well separated from the pump, thus there is no need of high-extinction notch filters required when SFWM is concerned \cite{orieuxRecentAdvancesIntegrated2016}. SPDC in lithium niobate has been widely used to generate highly entangled photon pairs and many experiments used periodically poled bulk crystals, titanium indiffused, proton exchanged or etched waveguides as sources of quantum states for quantum information applications  \cite{honjoGenerationEnergytimeEntangled2007, alibartQuantumPhotonicsTelecom2016, zhaoHighQualityEntangled2020, xueUltrabrightMultiplexedEnergyTimeEntangled2021}. Recent results have also shown that LNOI can be used to perform reliable quantum state generation, qubit control and single photon routing; building blocks for integrated quantum computing are thus being developed \cite{chapmanQuantumLogicalControlledNOT2023, maederOnchipTunableQuantum2024, chapmanOnchipQuantumInterference2024}. Furthermore, single-photon detectors integrated on LNOI circuits have been reported \cite{lomonteSinglephotonDetectionCryogenic2021, prencipeWavelengthSensitiveSuperconductingSinglePhoton2023}. As of today, however, there has been no demonstration of generation and reconstruction of time-bin entangled qubits on a single LNOI circuit, which is the natural evolution of a platform aiming at becoming leader of optical communication technologies. 
	
	Here, we present an integrated photonic circuit on LNOI that generates time-bin entangled Bell states by SPDC on a periodically-poled lithium-niobate (PPLN) waveguide with \SI{242}{\MHz/\mW} on-chip brightness, and performs its tomographic reconstruction with \SI{91.9}{}$\pm$\SI{1.0}{\%} fidelity without background subtraction or active phase stabilisation. We observe quantum interference with \SI{78.1}{}$\pm$\SI{2.0}{\%} visibility, limited only by chromatic dispersion, thus confirming that the generated states are beyond the limit of local hidden variables and can be exploited as a useful resource of quantum information. We observe three orders of magnitude higher source brightness when compared to similar experiments on silicon-based technologies \cite{xiongCompactReconfigurableSilicon2015,zhangIntegratedSiliconNitride2018}.
	
	\section*{Results}
	The designed optical circuit is illustrated in Fig.~\ref{fig:01_Device}(a). It consists of a single-mode waveguide circuit etched into a \SI{300}{\nm}-thick x-cut magnesium-oxide-doped lithium niobate film. Pulse pairs from an \SI{80}{\MHz} mode-locked laser at \SI{775}{\nm} are generated on the optical table by means of a fibre-based unbalanced interferometer. Before coupling into the device, pulses propagate across approximately \SI{15}{\m} of polarisation-maintaining optical fibre, and are thus stretched by chromatic dispersion from the original transform-limited pulse duration of \SI{100}{\fs} to an estimated of \SI{17}{\ps}. The time-bin delay between two pulses is $\tau \approx$~\SI{220}{\ps} to be able to easily resolve all time-bins with our single photon detectors. Pump light is input from the left, coupling in and out of the device is done by using focused grating couplers (GC). A wavelength-division multiplexing device allows to probe the circuit at visible or near infrared wavelengths by selecting the input with appropriate GC.
	The device consists of two twin Franson interferometers; the long arm is constituted of a delay line matching the pump pulses time separation, while the short arm is equipped with a variable optical attenuator (VOA), realised with a Mach-Zehnder interferometer, tuned by thermo-optic phase shifters (TOPSs) \cite{maederOnchipTunableQuantum2024}. VOAs are used to compensate for additional propagation losses experienced by the photons travelling on the long arm and maximise interferometric visibility at the recombination point. An additional TOPS running around the delay-line is used to tune the phase of signal and idler photons ($\varphi_s$, $\varphi_i$), thus implementing projections of the interfering photons.
	
	Setting up the experiment requires finely tuning the delay of the free space interferometer to match that of the fabricated circuit; this is done by pumping the system in reverse, with a femtosecond laser at \SI{1550}{\nm} which propagates backwards across the delay-line first, and after recombination pulses are fed into the \SI{2}{\mm}-long PPLN for second-harmonic generation (SHG). Simultaneously tuning the VOA to balance the pulses intensity and sweeping the applied voltage to the delay line shows interference of the up-converted signal at the preparation interferometer output. Interference is expected to manifest maximum \SI{50}{\%} visibility (four equally intense pulses are present in total, only two of which recombines in time while the other two constitute a constant background), as it is indeed observed and reported in Fig.~\ref{fig:01_Device}(b). Deviations from the sinusoidal behaviour are to be attributed to random phase fluctuations in the table-top interferometer, which is not actively stabilised.
	
	Next, we quantify the pump power required to minimise double down-conversion events following the procedure described in \cite{marcikicTimebinEntangledQubits2002} on a straight PPLN waveguide fabricated next to the device: the photon generation probability can be estimated by taking the ratio between coincidence counts at zero time delay and at the repetition period of the pump laser. We record coincidences for \SI{60}{\s} over a \SI{30}{\ns} window; the measurement outcome is displayed in Fig.~\ref{fig:01_Device}(c), fitted to a linear model as predicted by theory and shows that, at \SI{10}{\uW} of off-chip pump power, the probability of generating a single SPDC pair per pulse is $p=$~\SI{2.79}{}$\pm$\SI{0.09}{\%}, thus that of double events is proportional to $p^2\approx$~\SI{0.08}{\%} and can be neglected.  Errors concerning single photon measurements are estimated by assuming Poissonian statistics of the photon counts. Given the GC efficiency of $\sim$\SI{7}{\dB} and pump pulse duration, we estimate an on-chip average and peak pump power of \SI{2}{\uW} and \SI{1.5}{\mW}, respectively. At \SI{10}{\uW} off-chip pump power, we measure entangled states at $\sim$\SI{1800}{\Hz} which, considering the measured Klyshko efficiency \cite{klyshkoUtilizationVacuumFluctuations1977} of $\eta=$\SI{-15.5}{\dB} (see Supplementary Material), gives an on-chip source brightness of $\sim$\SI{242}{\MHz/\mW}. The measured values highlight the advantage of using LNOI and SPDC against the more common silicon-based platforms for quantum information applications, as we can achieve similar or higher rates by pumping our source with two to three orders of magnitude lower pump peak power compared to other, fully integrated devices based on SFWM \cite{xiongCompactReconfigurableSilicon2015, zhangIntegratedSiliconNitride2018}.
	
	At sufficiently low pump power, the down-converted photons at the PPLN create an energy-time entangled pair in the form of a Bell state that reads
	\begin{equation}
		\ket{\psi} = \frac{1}{\sqrt{2}} \left(\ket{EE}+e^{i\varphi_p}\ket{LL}\right),
	\end{equation}
	where $\varphi_p$ is the relative phase between pump pulses, control on which allows to generate the maximally entangled $\ket{\Phi^\pm}$ states. $\ket{E}$ ($\ket{L}$) describe whether the photon pair has been generated by the leading (trailing) pulse. Due to the type-0 SPDC process, we are forced to probabilistically split photons via Y-junctions; they are then sent to the analysis interferometers. After recombination, photons are out-coupled via infrared GCs and collected by v-groove fibre arrays and guided to commercial superconducting nanowire single-photon detectors; a time-tagger is synchronised to the pump laser clock for counts binning. Further device characterisation is available in the Supplementary Material.
	
	\begin{figure}[t!]
		\centering
		\includegraphics[width=0.499\textwidth]{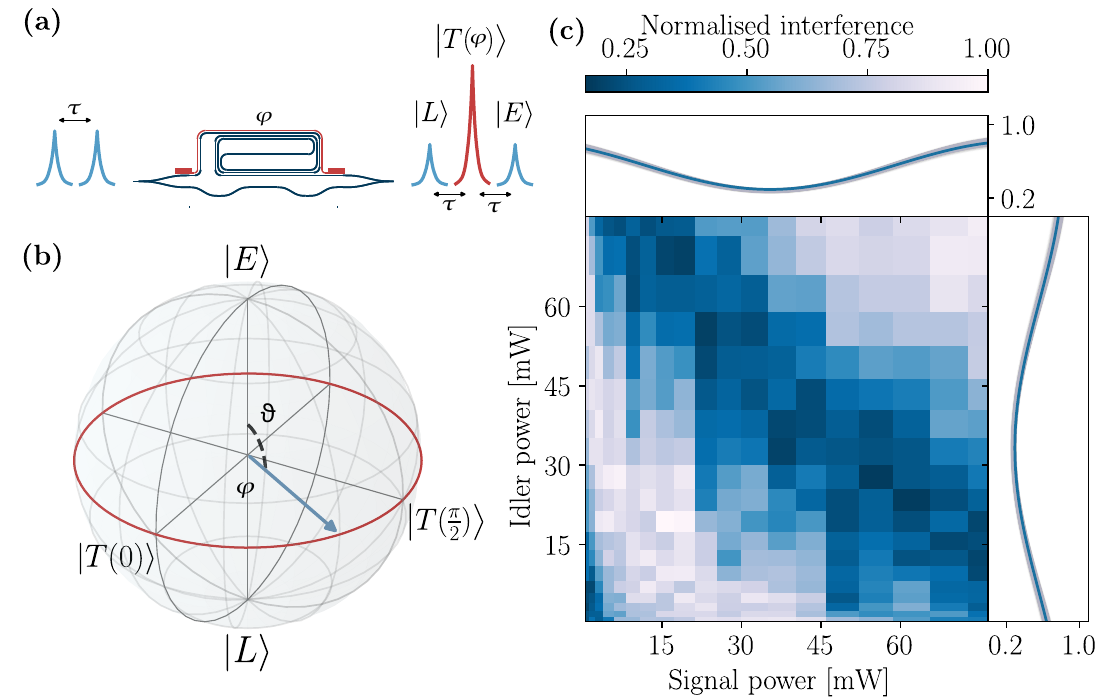}
		\caption{\textbf{Quantum interference calibration.} (a) Schematic of interfering state generation. (b) Single qubit Bloch sphere; the equator can be swept across by tuning the relative phase between interferometer arms. (c) Quantum interference calibration map: oscillation of the quantum state projected onto the energy basis over a wide enough range to span the full Bloch spheres equator. Top/side insets show the interference trend for the signal and idler channels, respectively. Uneven sampling of the map is due to the measurement being taken in voltage steps rather than power.}
		\label{fig:02_Calibration}
	\end{figure}
	
	\begin{figure*}[t!]
		\centering
		\includegraphics[width=0.8\textwidth]{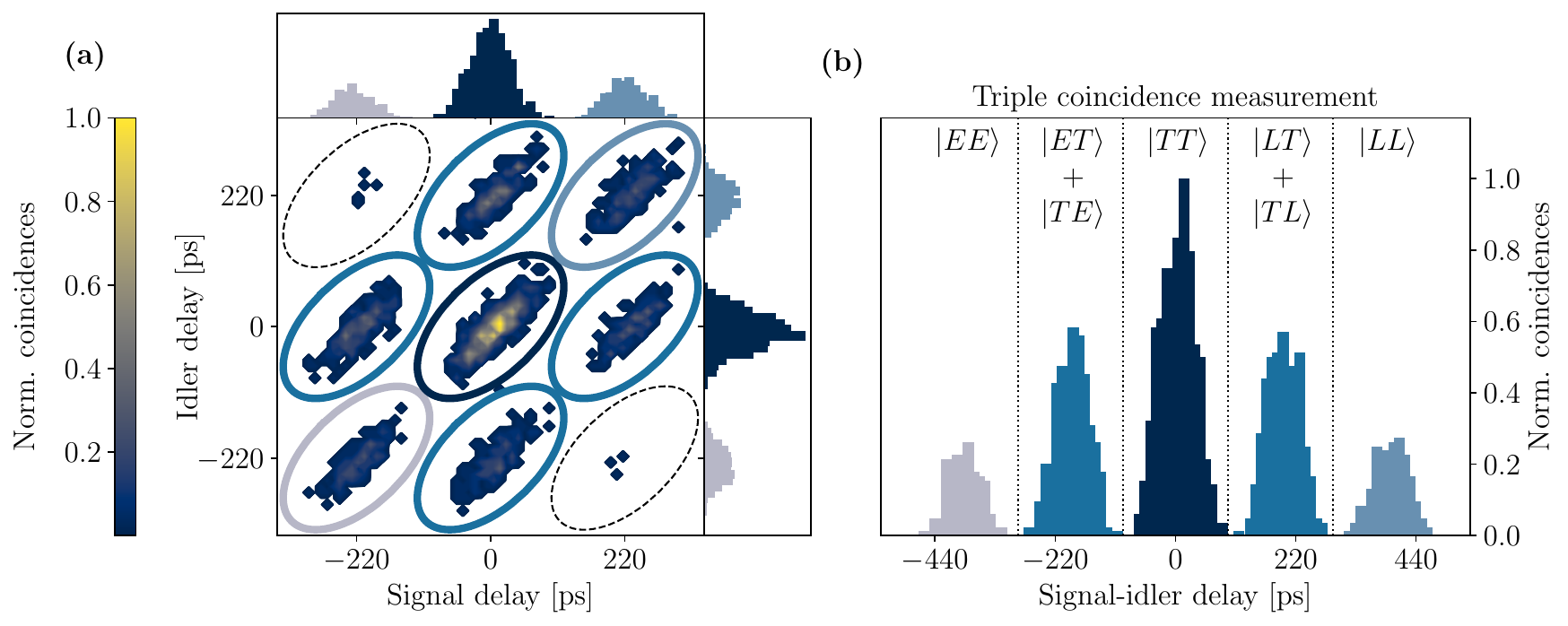}
		\caption{\textbf{Triple coincidence measurement.} (a) Two-dimensional coincidence histogram between signal, idler and pump trigger. Insets show coincidence counts relative to the pump clock on signal and idler channels independently. Drawn ellipses highlight the possible measurement outcomes and are colour-coded correspondingly to the peaks of the collapsed, one-dimensional histogram in (b). Labels in panel (b) explicitly mark which measurement outcome corresponds to each histogram peak, with the central one representing the interfering on-time state.}
		\label{fig:03_Stars}
	\end{figure*}
	
	Generation of the interfering state and its representation on the qubit Bloch sphere are illustrated in Fig.~\ref{fig:02_Calibration}(a-b). Before entering the analysis interferometer, qubits are in a superposition of $\ket{E}$ and $\ket{L}$ states, which constitute the so-called \emph{time basis} of our experiment and lay at the qubit Bloch sphere poles. After traversing the interferometer, when the early (late) photon has propagated across the long (short) path, an additional state arises as there is no possibility of inferring whether is the early photon to have accumulated a delay or the late photon to have been advanced by a time-bin. This gives rise to quantum interference, which can be controlled by tuning the relative phase between the two interferometer arms. We refer to this additional state as \emph{on-time} state, labelled $\ket{T(\varphi)}$, and its probability amplitude oscillates as a function of $\varphi = \varphi_s+\varphi_i-\varphi_p$. $\ket{T}$ lies on the Bloch sphere equator and specific values of the relative phase allow to build the so-called \emph{energy basis}, which being constituted of non-orthogonal eigenvectors to the time basis allows to perform tomographic reconstruction and to validate the presence of entanglement. After propagation through the interferometer, early and late qubits accumulate twice the delay, we avoid re-defining them for the sake of keeping a light notation. Early, late and on-time state form a \emph{qutrit} to be analysed by Franson interferometry. 
	
	We calibrate quantum interference by sweeping the electrical power applied to signal and idler TOPSs, and record the oscillation of coincidence counts corresponding to the $\ket{TT}$ state to identify the thermo-optic powers  related to different projectors. We perform triple coincidence measurements (signal, idler and pump trigger signal) and collect statistics about photon arrivals in the interfering state. Figure~\ref{fig:02_Calibration}(c) shows a two-dimensional calibration map of signal-idler coincidences with respect to the trigger signal, as a function of signal and idler TOPS powers. Experimental data are fitted to a sinusoidal model and the resulting interference behaviour is shown in the Fig.~\ref{fig:02_Calibration}(c) top/side insets separately for signal and idler channel, respectively. Shaded confidence regions obtained by applying Poissonian statistics are also reported. The map of Fig.~\ref{fig:02_Calibration}(c) covers a wide-enough region to sweep the two projectors across the entire Bloch sphere equator for the two qubits. It serves both to identify the required voltages, but also to extract the pump phase and rotate the reference frame in order to reconstruct a maximally entangled Bell state by the simple transformation
	$\tilde{\varphi}_{s,i} \rightarrow \varphi_{s,i}-\varphi_p/2$.
	
	\begin{figure*}[t!]
		\centering
		\includegraphics[width=0.8\textwidth]{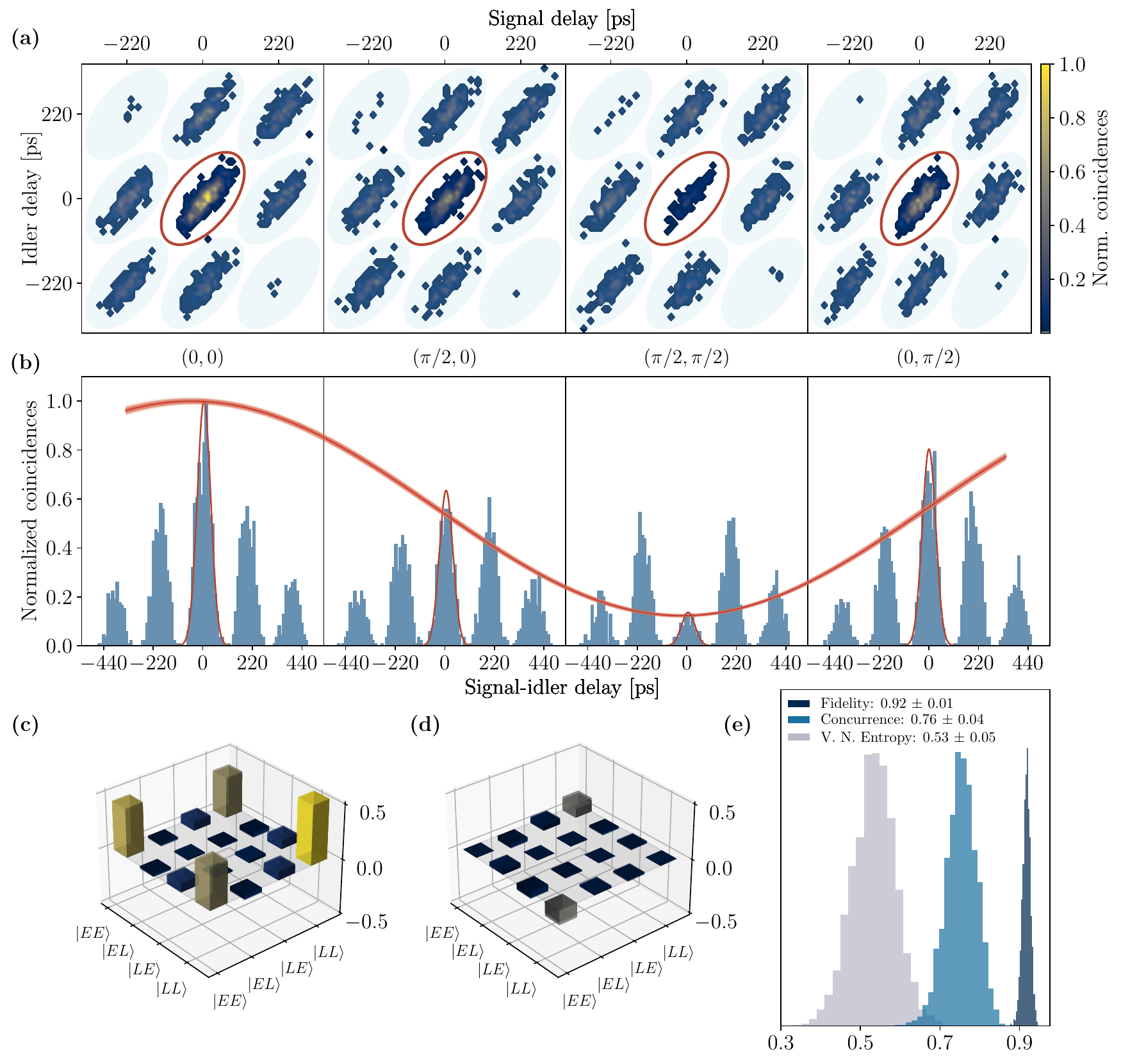}
		\caption{\textbf{Observed quantum interference and reconstructed density matrix.} (a) Two-dimensional histograms correspondent to the four sets of projectors, with the oscillating state marked with a red ellipse. (b) Collapsed triple coincidences with the interfering state fitted to a Gaussian model and the corresponding oscillatory behaviour highlighted. (c-d) Tomographic state reconstruction, real and imaginary part, with shaded bars superimposed to the matrices indicating errors obtained via Monte Carlo simulation. (e) Fidelity, concurrence and von Neumann entropy histograms originating from the Monte-Carlo simulation, showing high degree of entanglement and suitability of the device for QKD experiments.}
		\label{fig:04_Interference}
	\end{figure*}
	
	Triple coincidence measurements between the common pump clock and photons at the two channels are at the core of our experimental procedure. They enable, thanks to the characteristic feature of Franson interferometry, to simultaneously measure in two non-orthogonal bases, thus the observation of all the possible quantum states originating from probabilistic photon splitting. At the interferometer output and before detection, the quantum state in the joint space, $\ket{\tilde{\psi}(\varphi)}$, features seven possible outcomes (for more information, see Supplementary Material). These can be visualised with a two-dimensional histogram reporting coincidences between signal, idler and pump trigger. This novel measurement technique fully exploits Franson interferometry and increases the amount of retrievable information. An example histogram is illustrated in Fig.~\ref{fig:03_Stars}(a), where nine elliptical patterns develop as a function of signal and idler delays, and correspond to each of the possible measurement outcomes. Along the principal diagonal, we find states $\ket{EE}$, $\ket{TT}$ and $\ket{LL}$, which are those of interest for entanglement-based time-bin encoding QKD systems. Off the diagonal are non-interfering states correspondent to distinguishable events to which only early or late photons contribute by taking opposite paths on the two interferometers. The remaining two patterns at the extremes of the anti-diagonal, marked with black dashed ellipses, correspond to outcomes that should not occur if the purest Bell state is produced and we attribute to residual double down-conversion events. Top and side insets show projections of the recorded coincidence on signal and idler channel, respectively, and correspond to the typical measurement that would be performed by each of the two communicating parties. Figure~\ref{fig:03_Stars}(b) shows a collapsed view of the triple coincidence measurement outcome in one dimension, which features five peaks as reported by other works \cite{jayakumarTimebinEntangledPhotons2014, chenInvitedArticleTimebin2018}. The drawn ellipses in panel (a) are colour-coded correspondingly to the peaks to which counts belong in the collapsed view of panel (b), and labels indicate which specific state falls within each of the histogram sections. It is important to mention that deterministic photon routing would enable the a-priori selection of the quantum state to be observed and eliminate the spurious states belonging to second and fourth histogram peaks, which do not carry useful information. This would allow to optimise data collection and increase coincidence rates depending on whether security of the QKD link is to be confirmed by entanglement quantification or information sharing is desired.
	
	\begin{figure*}[t!]
		\centering
		\includegraphics[width=0.9\textwidth]{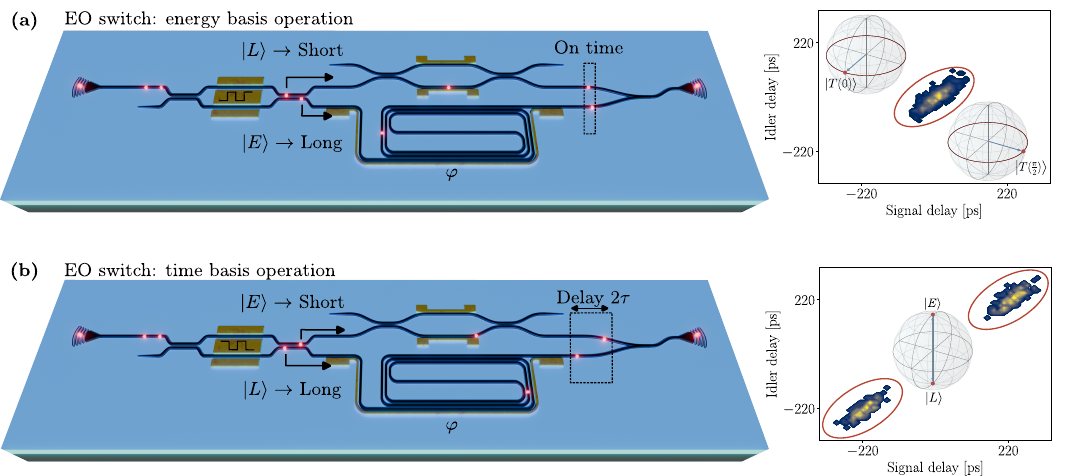}
		\caption{\textbf{Proposed QKD receiver device.} Electro-optic modulated Franson interferometer for deterministic photon routing. (a) Energy basis regime to maximise counts flowing into the interfering state and confirm channel security. (b) Time basis regime to deterministically split qubits in time and use them as source of secure quantum information. Example two dimensional histograms and state representation in the Bloch sphere are reported on the side panels for the two cases.}
		\label{fig:05_EOdevice}
	\end{figure*}
	
	\subsection*{Quantum state tomography}
	Entanglement quantification between the down-converted photons is performed by quantum interference measurement. We set electrical powers on the two interferometers corresponding to projectors at $(\varphi_s, \varphi_i )= (0,0), (\pi/2,0), (\pi/2, \pi/2)$ and $(0,\pi/2)$ and record coincidences for \SI{300}{\s} at each point. Counts belonging to the interfering state are observed to oscillate with \SI{78.1}{}$\pm$\SI{2.0}{\%} visibility without need of background subtraction. Figure~\ref{fig:04_Interference}(a-b) displays the measurement results: two-dimensional histograms (a) and collapsed triple-coincidence measurement (b) for each set of projectors. Counts pertaining to the interfering state are fitted to a Gaussian model, which is integrated over five standard deviations. The integrated counts are used as fitting points to extract the sinusoidal behaviour, with the probability of detecting the interfering state that features the expected oscillation according to
	
	\begin{equation}
		P = |\bra{TT}\tilde{\psi}\rangle|^2 \propto 1+\cos\left(\tilde{\varphi}_s+\tilde{\varphi}_i\right).
	\end{equation}
	
	We measure an average coincidence rate of \SI{5}{\Hz}, limited by the pump laser repetition rate, additional losses introduced by the time-bin decoding device (probabilistic splitting \SI{3}{\dB}, propagation loss \SI{1}{\dB}, estimated at the VOA output), and the need of further filtering one of the two channels ($>$\SI{10}{\dB}). The broad type-0 SPDC photon bandwidth indeed forces us to place an additional \SI{8.8}{\nm} (full-width at half-maximum) bandpass filter on signal or idler channels, otherwise chromatic dispersion in the waveguide would heavily impact visibility due to the different group velocities of non-degenerate photon pairs. We filter only one channel and rely on energy conservation to post-select photons within a narrower spectral range in order to increase visibility. We measure an interference visibility at the limit of what the filtered photon bandwidth allows; an extended discussion on effects impacting visibility is available in the Supplementary Material. Photon bandwidth and deterministic splitting can be simultaneously achieved by adopting type-II \cite{kimStudyTypeII2021} or counter-propagating \cite{luoCounterpropagatingPhotonPair2020} SPDC. These approaches would immediately yield more than \SI{13}{\dB} increase in coincidence counts without changing experimental setup or procedure, while additional \SI{11}{\dB} could be gained by increasing the pump repetition rate to \SI{1}{\GHz}. Importantly, the average pump power used in this experiment is highly conservative as it corresponds to a probability of double down-conversions lower than \SI{0.1}{\%}. We chose this regime in order to avoid further limiting factors to the visibility other than chromatic dispersion. The pump power could easily be doubled with a corresponding two-fold increase in photon counts with better engineered dispersion or photon bandwidth. 
	
	We then perform quantum state tomography of the generated qubits following the procedure described in \cite{jamesMeasurementQubits2001, takesueImplementationQuantumState2009}. Maximum likelihood estimation allows us to reconstruct the quantum state with \SI{91.9}{}$\pm$\SI{1.0}{\%} fidelity to a Bell state $\ket{\Phi^+}$. Real and imaginary parts of the density matrix are displayed in Fig.~\ref{fig:04_Interference}(c-d), with errors being illustrated by shaded bars on top of the reconstructed matrix and calculated by running five thousand iterations of a Monte-Carlo simulation assuming Poissonian statistics of photon counts. We observe a residual phase in the reconstructed matrix, which we attribute to pump phase fluctuations in the preparation interferometer that is not actively stabilised. This nevertheless shows that our calibration procedure is robust against small phase fluctuations and allows us to immediately set the desired projection angles. Figure~\ref{fig:04_Interference}(e) shows histograms of the calculated relevant quantities on the reconstructed quantum state obtained via Monte-Carlo simulation. Other than the fidelity, we report a concurrence of \SI{0.76}{}$\pm$\SI{0.04}{}, which is enough to guarantee violation of the Clauser-Horne-Shimony-Holt inequality and further proves the presence of highly entangled qubits \cite{clauserProposedExperimentTest1969, verstraeteEntanglementBellViolations2002}. We quantify the von Neumann entropy of the generated state to \SI{0.53}{}$\pm$\SI{0.05}{\bit}, which is close to the expected value of \SI{0.5}{\bit} characteristic of a maximally entangled Bell state. These results confirm the suitability of our platform as transmission and reception modules of quantum information for entangled time-bin encoding QKD protocols.
	
	Last, we propose a modification to our device towards an implementation as QKD receiving module. Addition of high-speed EO switches would enable to route early and late photons to the desired interferometer arm for projection in the time or energy basis, as illustrated in Fig.~\ref{fig:05_EOdevice}. In the prospect of employing the device for QKD demonstrations, the operation regime would be alternated in order to periodically confirm the presence of entanglement by directing all photons to the interfering state ($\ket{E}$ to the long arm and $\ket{L}$ to the short arm) (Fig.~\ref{fig:05_EOdevice}(a)) in order to maximise counts in the central peak and rapidly confirm security of the link. Once entanglement is confirmed, the operation can be reversed in order to temporally split the photons which would then serve as unconditionally secure source of information by leveraging entanglement (Fig.~\ref{fig:05_EOdevice}(b)). The side panels display the corresponding pattern that would develop on the two-dimensional histogram for the two operation regimes, and projected states are highlighted in the accompanying Bloch spheres. One option to realise QKD modules would be to have a first device on Alice's side performing state preparation and projection measurement of one qubit. The second qubit would be sent to Bob's receiver chip on the opposite side of the link, and he would then perform projection measurements on it. The communication scheme would then follow the typical procedure of E91 or BBM92 protocols. Incidentally, with the described conditions, state preparation could also be performed by a third party, Charlie, who would distribute the two qubits to the communicating parties in a semi-device-independent scheme.

	\section*{Discussion}
	Integrated photonic circuits are a promising solution to the long sought implementation of large-scale, unconditionally secure quantum communication networks. In this work, we showed that lithium niobate-on-insulator, that has been already proved leading platform for classical telecommunication, can be exploited also in the quantum regime. Our results show that thin-film lithium niobate outperforms silicon-based technologies in the generation and reconstruction of entangled quantum states in terms of state generation efficiency, thus required pump energy to achieve similar or higher single photon count rates. By combining our results with the already available high-speed switches and integrated single photon detectors, there is clear potential for the development of reliable transmitters and receivers of quantum information on a single platform to enhance communication security using standard telecommunication networks. While at the current stage information rates are still limited by the pump laser repetition and the need of photon filtering, pulses can be carved out of continuous wave laser diodes, heterogeneously integrated on the platform \cite{shams-ansariElectricallyPumpedLaser2022}, in order to boost the data-rate. We chose \SI{220}{\ps}-wide bins due to a conservative approach, but the limiting factor of our system is only its timing jitter, which corresponds to approximately \SI{50}{\ps}. We estimate that pulses could be carved at \SI{10}{\GHz}, which corresponds to \SI{100}{\ps} pulse pairs at a \SI{5}{\GHz} rate. Under these conditions, by employing deterministic SPDC photon separation, reducing their bandwidth or engineering waveguide dispersion would yield more than \SI{40}{\dB} in coincidence count-rates (\SI{18}{\dB} for increased pump rate, \SI{3}{\dB} for double pump power, \SI{10}{\dB} for removing the need of filtering, \SI{8}{\dB} for EO modulation, \SI{3}{\dB} for deterministic separation of SPDC). Additionally, lower losses between circuit and detection would also translate into an increase in count-rates, with the ultimate goal being integration of single photon detectors on chip. Moreover, the full bandwidth can be exploited to wavelength-multiplex the communication and further increase channel capacity \cite{xueUltrabrightMultiplexedEnergyTimeEntangled2021, muellerHighrateMultiplexedEntanglement2024} towards \SI{}{\MHz}-rate entangled quantum key distribution. Our results confirm that lithium niobate-on-insulator is a promising platform for energy-saving solutions to the quest of realising integrated communication modules for large scale quantum information networks towards unconditionally secure data sharing.
	
	{\small 
		\section*{Methods}
		\subsection*{Design and fabrication}
		Optical circuits and delay lines are simulated using finite element simulations and custom scripts to extract chromatic dispersion in the long spirals by taking into account anisotropy of the crystal. VOAs and TOPSs are designed by combining optical finite element with heat transfer simulations and a coupled-mode theory approach as in \cite{maederOnchipTunableQuantum2024}. The devices are fabricated on a \SI{300}{\nm}-thick, \SI{5}{\%} MgO-doped, x-cut lithium niobate-on-insulator substrates with \SI{2}{\um}-thick buried oxide layer on a silicon handle. First, comb-shaped electrodes for periodic poling are deposited by patterning with electron-beam lithography (EBL) a double-layer polymethyl methacrylate (PMMA) resist, followed by lift-off of \SI{100}{\nm} of chromium. The crystal is poled by applying a train of high-voltage pulses and poling quality is ensured by non-invasive two-photon microscope imaging. An array of PPLN regions is patterned on the sample at first, and after imaging the layout is adapted to connect those best matching the design parameters to the Franson interferometers for the next lithography step. Waveguides are defined by EBL on a \SI{500}{\nm}-thick hydrogen silsesquioxane resist and the film is etched \SI{200}{\nm} by Argon ion milling. Redeposited material is removed by wet cleaning using a potassium hydroxide solution following the process described in \cite{kaufmannRedepositionfreeInductivelycoupledPlasma2023}. TOPSs are patterned with an additional EBL step and lift-off process of \SI{100}{\nm} of gold with \SI{5}{\nm} of chromium adhesion layer. Electrodes are routed to the edge of the chip by means of a second gold layer patterned by direct laser writing and lift-off. Finally, electrodes are wire-bonded to a printed circuit board for simultaneous control of all heaters via software.
		
		\subsection*{Measurements and data analysis}
		A Ti:Sapphire mode-locked laser at \SI{80}{\MHz} provides pump pulses for SPDC. The laser is coupled into polarisation-maintaining fibre and guided to a fibre-based table-top unbalanced interferometer where pulse pairs are produced. Pulses propagate for approximately \SI{15}{\m} of fibres, which according to the manufacturer's data manifest a chromatic dispersion parameter of roughly $D=$\SI{-130}{\ps/\nm/\km} at the pump wavelength. Given the pump bandwidth of \SI{8.8}{\nm} at the source, we calculate a temporal broadening of $\sim$\SI{17}{\ps} before coupling into the chip. Light is coupled in and out of the devices by using v-groove fibre arrays through focused grating couplers etched into the LN film. Classical characterisation is possible by either pumping the system in reverse and inspecting the generated second harmonic signal from a femtosecond telecom laser, or forward by using a wavelength division multiplexing device that allows to couple \SI{1550}{\nm} light into the circuit by evanescent field coupling. Additional GCs at all ports of the Mach-Zehnder interferometers (not illustrated in the circuit schematic) allow to cross-check the device behaviour at all configurations of input/output. Custom python software, a voltage source and custom electrical circuitries allow to automatise all classical and quantum measurements. VOAs are characterised by monitoring the additional ports and \SI{1550}{\nm} pulses are measured with a fast photodiode and oscilloscope while the tuning voltage is swept until the balance point is found. Balancing the interferometers can also be done by monitoring single photon coincidences and making sure that histograms corresponding to $\ket{E}$ $\ket{L}$ states have equal amplitude. Time-stamping single photon counts is done by synchronising our time-tagger device with an electrical trigger signal extracted from the pump laser cavity. For single photon measurements, device outputs are filtered with long-pass filters ($>$\SI{60}{\dB} extinction) on both channels to remove residual pump photons that would saturate the detectors or add deleterious background. An additional bandpass filter on the idler channel allows to post-select photons within a narrow spectral region for enhanced interference visibility. 
		
		Data is analysed in Python with custom software partially based on the QuTip package. 
		
		\subsection*{Acknowledgements}
		We acknowledge support for characterization of our samples from the Scientific Center of Optical and Electron Microscopy ScopeM and from the cleanroom facilities BRNC and FIRST of ETH Zurich and IBM Ruschlikon. R.G. acknowledges support from the European Space Agency (Project Number 4000137426), the Swiss National Science Foundation under the Bridge program (Project Number 194693) and the Swiss Consolidator (Project Number 2022 213713). R.J.C. acknowledges support from the Swiss National Science Foundation under the Ambizione Fellowship Program (Project Number 208707).
		
		\subsection*{Competing interest}
		The authors declare no competing financial or non-financial interests.
		
		\subsection*{Data availability statement}
		Data supporting the findings of this study are available within the article and Supplementary Material. Raw data and analysis code are available from the corresponding author upon reasonable request.
		
		\subsection*{Author contributions}
		G.F. designed and fabricated the device, performed the experiments and data analysis, wrote the manuscript. F.M. developed and tested the building blocks, wrote the experiment automation routines. A.M. developed simulations routines and testing of VOAs and thermo-optic shifters, designed and set-up the electronics controlling the experiment. A.M., J.K. and R.J.C wrote measurement software for components characterisation. G.F., J.K. and R.J.C. performed classical characterisation of the device. J.K. and A.S. developed, optimised and performed periodic poling of the sample. R.J.C. and G.F. conceived the idea of the project. R.G. and R.J.C. supervised the project. All authors contributed to revising and validating the manuscript content.
		
	}

	\clearpage
	
	\renewcommand{\figurename}{Figure} 
	\renewcommand\thefigure{SF\arabic{figure}}    
	
	\onecolumngrid
	\begin{center}
		\LARGE \textbf{Supplementary Material}\\[2em]
	\end{center}

	\section*{SM1: Setup and characterisation}\label{sec:Sup01_device}
	\begin{figure}[h!]
		\includegraphics[width=\textwidth]{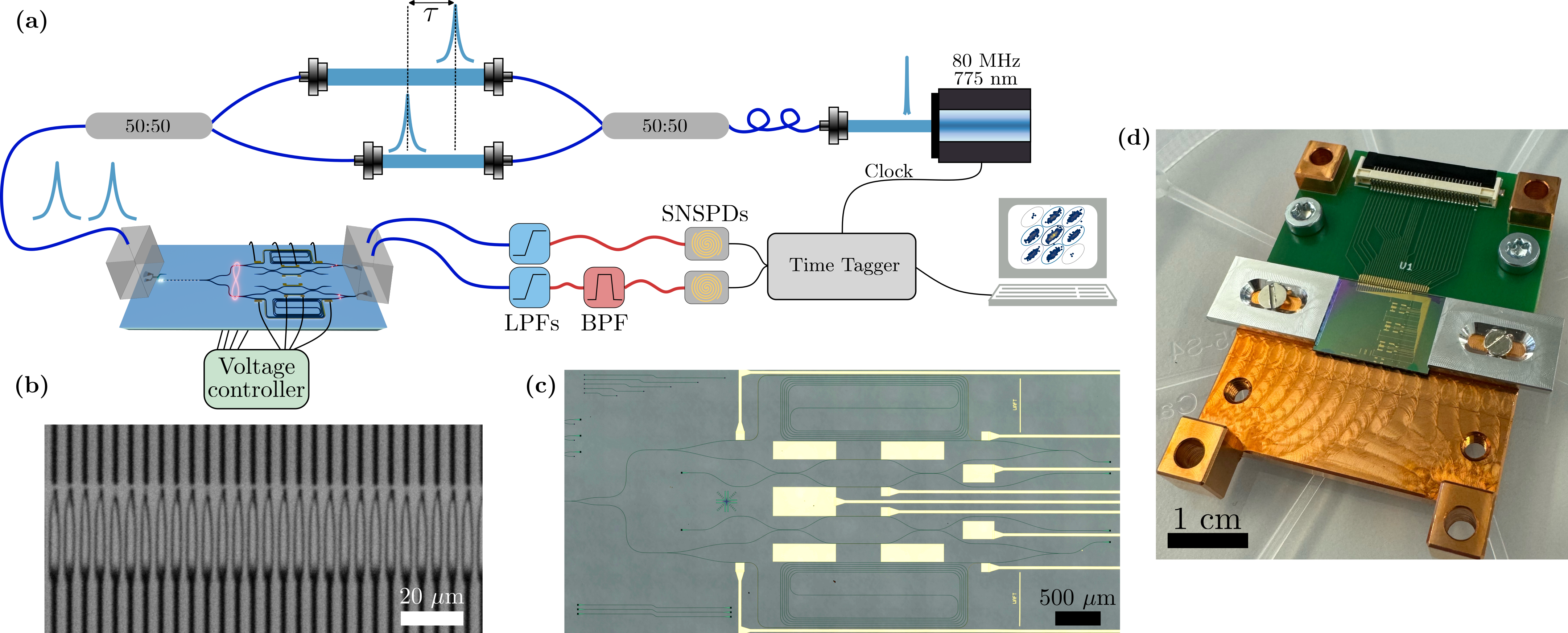}
		\caption{\textbf{Experimental setup and device.} (a) Schematic of the experimental setup. LPF: long-pass filter, BPF: bandpass filter, SNSPD: superconducting nanowire single photon detector.   (b) Two-photon microscope image of a fraction of poled waveguide, dark lines are the electrodes fingers. (c) Optical microscope image of a fabricated device, excluding input and poled waveguide. (d) Image of the final sample, including holder and printed circuit board.}
		\label{fig:S01_dev}
	\end{figure}
	
	Figure~\ref{fig:S01_dev}(a) illustrates a schematic of the experimental setup: an \SI{80}{\MHz} pump laser at \SI{775}{\nm} with \SI{100}{\fs} transform limited pulse duration is coupled into polarisation maintaining fibres and pulse pairs are generated with an unbalanced table-top interferometer. After propagating through approximately \SI{15}{\m} of polarisation maintaining optical fibre, pulses are stretched to an estimated duration of \SI{17}{\ps} given the dispersion parameter of $D = $\SI{-130}{\ps/\nm/\km} at \SI{775}{\nm}. Pulses and output photons are coupled in and out of the chip via v-groove fibre arrays and focused grating couplers. Two long-pass filters suppress the residual pump power and a bandpass filter on one of the two channels post-selects SPDC photons within a spectral range that allows to observe large interferometric visibility. Photon are detected with superconducting nanowire single photon detectors (SNSPDs) and time-binning is performed with a time-tagger system, synchronised with an electronic trigger signal from the pump laser. Figure~\ref{fig:S01_dev}(b) display a two-photon microscope image of a poled film region before patterning and etching of the waveguides. Figure~\ref{fig:S01_dev}(c-d) show an optical microscope picture of the device, excluding inputs and poled waveguide and a picture of the completed sample (which includes three copies of the device), respectively. 
	
\begin{figure}[h!]
		\includegraphics[width=\textwidth]{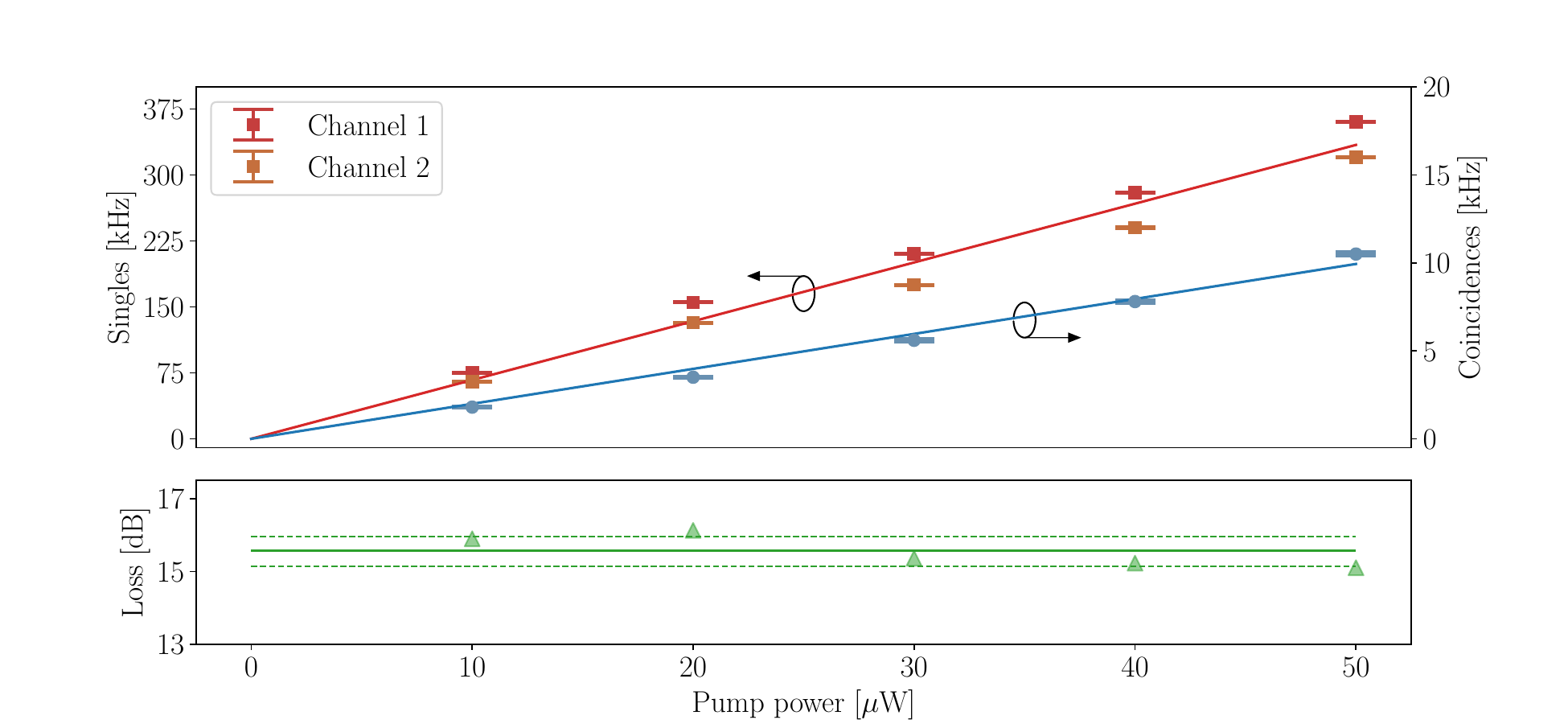}
		\caption{\textbf{Total system loss.} The top panel shows single counts at two SNSPDs channels and the measured coincidences along with linear fits to the data. Counts are generated and collected from a straight waveguide fabricated next to the device. The bottom panel reports the system loss versus off-chip pump power, with its average and standard deviation marked with horizontal solid and dashed lines, respectively.}
		\label{fig:S02_klyshko}
	\end{figure}

	We characterise source brightness and system loss by measuring the Klyshko efficiency \cite{klyshkoUtilizationVacuumFluctuations1977} on a straight waveguide fabricated next to the test device for time-bin experiments. The waveguide is pumped with varying average power, the output is collected and filtered to suppress the residual pump photons before being probabilistically split with a 50:50 fibre splitter. The two channels are monitored using SNSPDs to record both single and coincidence counts. The measured values are reported in Fig.~\ref{fig:S02_klyshko}, where the top panel shows single counts (for both channels) and coincidences versus off-chip pump power. Average singles and coincidences are fitted to a linear model as theory predicts. The bottom panel shows the system loss, calculated as the ratio between singles and coincidences, with the solid and dashed lines representing average loss and standard deviation, respectively. We measure \SI{15.5}{\dB} of average loss which gives, considering the grating coupler efficiency of $\sim$\SI{7}{\dB}, an on-chip source brightness of $\sim$\SI{242}{\MHz/\mW}. Additional \SI{7}{\dB} loss characterise the time-bin device due to probabilistic photon splitting and propagation loss in the waveguides.
	
\begin{figure}[h!]
		\includegraphics[width=\textwidth]{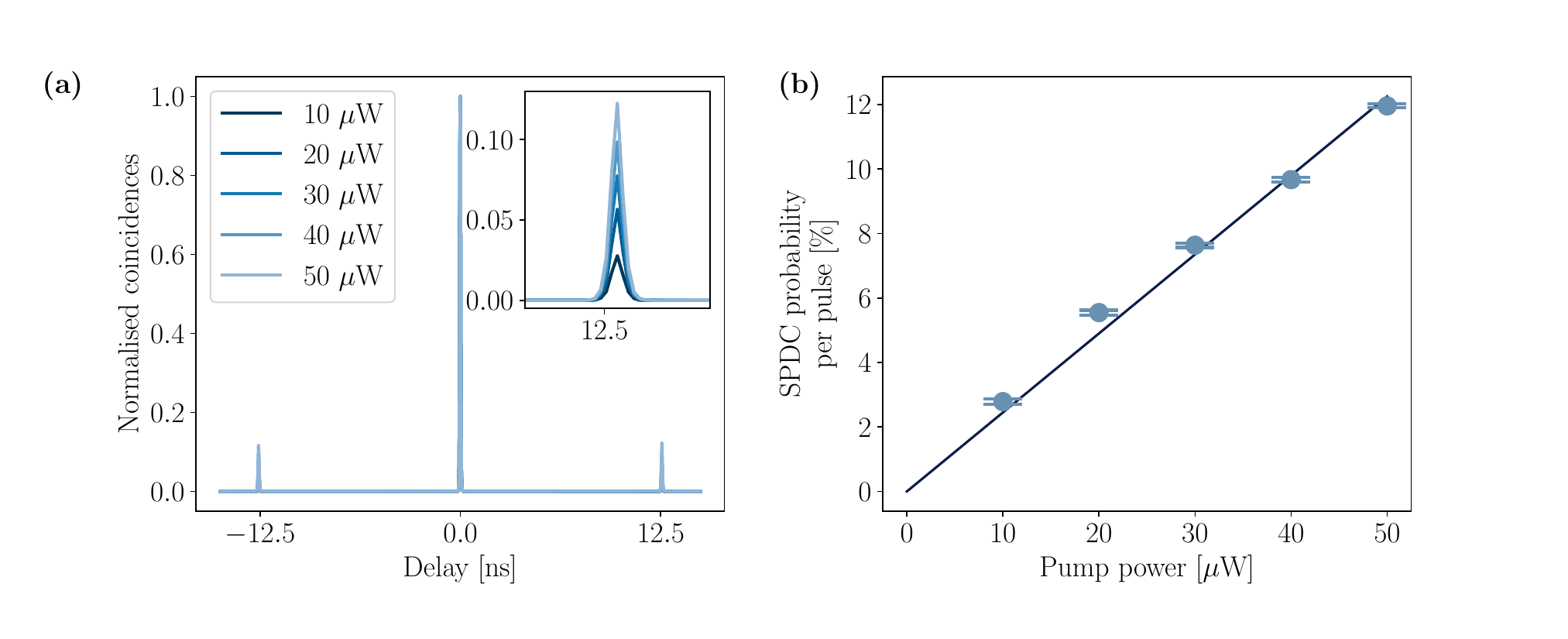}
		\caption{\textbf{Spontaneous parametric down-conversion probability.} Extended Figure 1(a) of the main manuscript, showing in panel (a) normalised coincidences over a time interval covering two pump periods, while the inset illustrates the secondary peak and its decrease in relative amplitude as the pump power is lowered. (b) SPDC generation probability per pulse versus off-chip pump power.}
		\label{fig:S03_spdc}
	\end{figure}

	We follow the procedure described in Ref.~\cite{marcikicTimebinEntangledQubits2002} to quantify the relationship between pump power and SPDC probability per pulse. Figure~\ref{fig:S03_spdc} displays the results and extends the data reported in the main manuscript. By calculating the ratio between coincidence peaks at zero time-delay at the repetition period of the pump laser, one can extract the probability of generating a photon pair from a pump pulse under the assumption of large enough system loss, which applies in our case. The workflow is based on the argument that if a photon from a pair is lost before detection, it may coincide with one originating from the next pulse in the sequence; this event is progressively more likely as the pump power is increased as the number of double down-conversion events is enhanced. Figure~\ref{fig:S03_spdc}(a) shows normalised coincidence counts over a time interval covering two pump periods, while the insets highlights the relative decrease of the secondary peak as the pump power is lowered, indicating a decreased probability of photon pair generation. SPDC generation probability per pulse versus off-chip pump power is reported again in Fig.~\ref{fig:S03_spdc}(b) 
	
	\begin{figure}[h!]
		\includegraphics[width=0.91\textwidth]{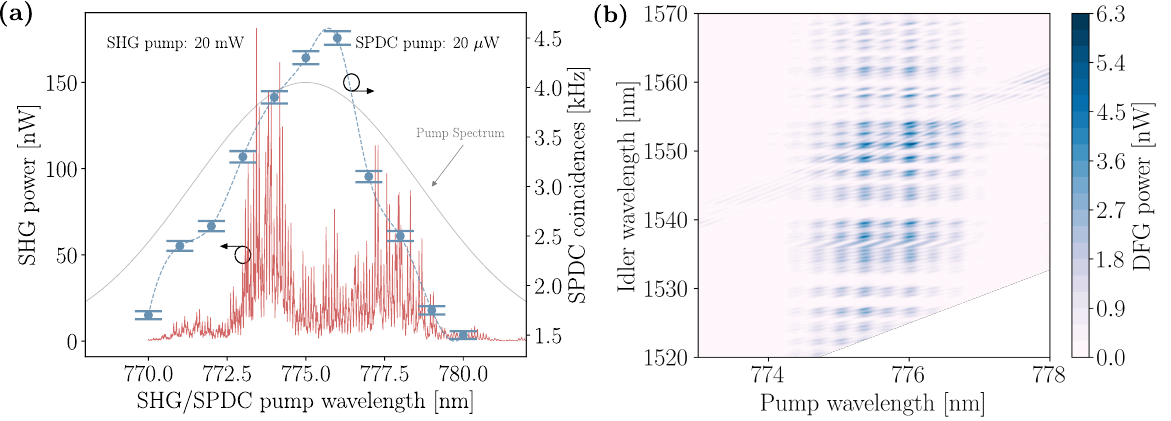}
		\caption{\textbf{Characterisation of periodically poled waveguides.} (a) Second harmonic generation spectrum and measured coincidence counts as a function of the pump wavelength. The thin solid envelope qualitatively illustrates the spectral width of the pump laser when centred at \SI{775}{\nm}. (b) Inferred difference frequency generation map illustrating the broad phase matching bandwidth of spontaneous parametric down-conversion in our waveguides.}
		\label{fig:S04_shg_spdc_sfg}
	\end{figure}
	
	Figure~\ref{fig:S04_shg_spdc_sfg} reports spectral characterisation data of our periodic poling. Figure~\ref{fig:S04_shg_spdc_sfg}(a) displays the second harmonic generation (SHG) spectrum (solid line) and measured coincidence counts (data-points, interpolated without specific functional form) for a straight waveguide fabricated next to a time-bin encoding device. The SHG spectrum is obtained by sweeping a continuous-wave infrared laser and recording the output SH power. Interference fringes originate from Fabry-Perot interference between small reflections at the input and output grating couplers. The relatively broad conversion bandwidth and skewed overall shape are due to the limited poled region length and thickness variations across the lithium niobate film, which shift the phase matching wavelength. Coincidence counts are recorded by tuning the central wavelength of the pulsed pump laser across the SH bandwidth. Although the pump laser spectrum (approximately \SI{9}{\nm}-wide, qualitatively displayed with a thin solid line in the background when centred at \SI{775}{\nm}) largely overlaps with the full SH bandwidth, we observe a consistent increase of coincidence counts as the central wavelength is swept across it in correspondence of the most efficient SH generation. Figure~\ref{fig:S04_shg_spdc_sfg}(b) illustrates an inferred difference-frequency generation (DFG) map. We conduct sum frequency generation measurements by sweeping two continuous-wave telecom lasers across the phase matching bandwidth and reconstruct the expected DFG map from it by energy conservation arguments. The results highlight the broad phase matching bandwidth of more than \SI{50}{\nm} by relating pump wavelength on the x-axis and idler (or signal) wavelengths on the y-axis. This confirms that the generated SPDC photons can be highly non-degenerate and forces us to utilise a bandpass filter to post-select photons within a narrower spectral range in order to enhance visibility as already discussed.
	
	\newpage
	\section*{SM2: State description and device calibration}\label{sec:Sup02_calib}
\begin{figure}[h!]
		\includegraphics[width=0.9\textwidth]{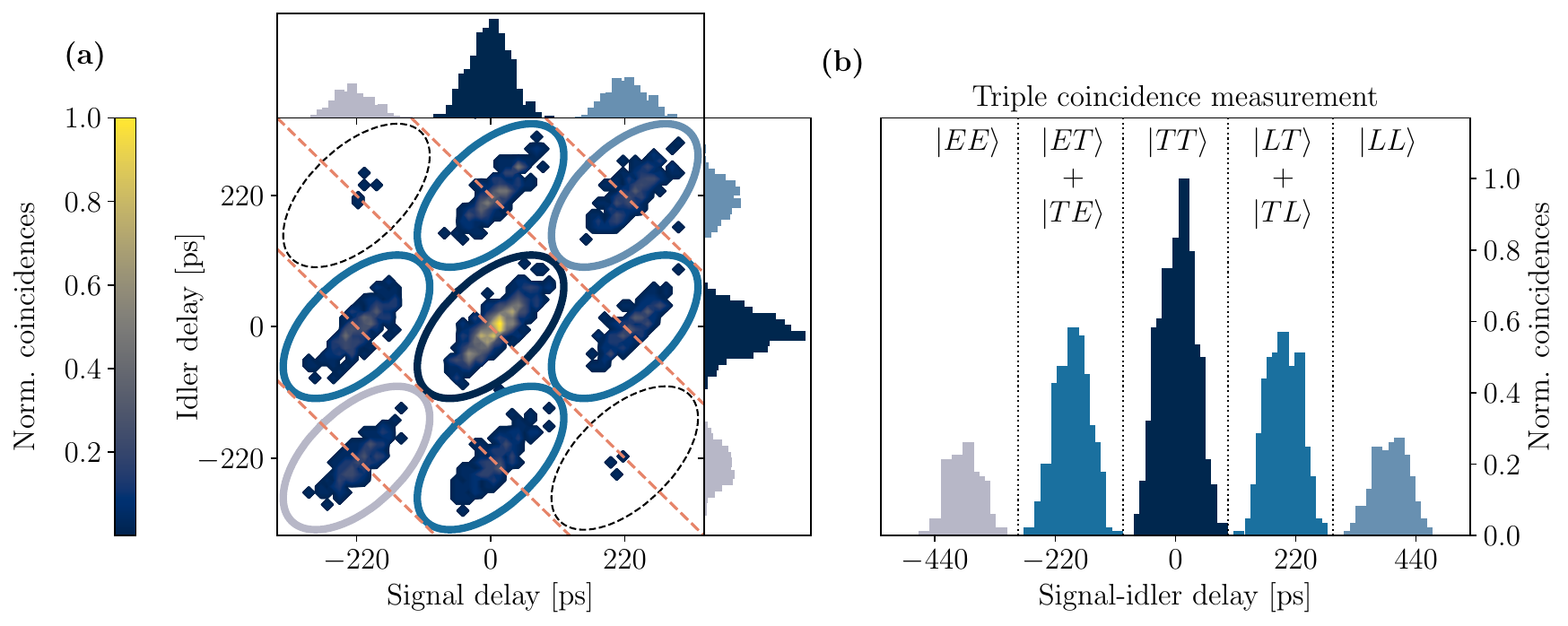}
		\caption{\textbf{Extended Figure 3: Triple coincidence measurement.} (a) Diagonal lines are superimposed to the two-dimensional histogram in order to better highlight the correspondence between each state and its representation in the collapsed view of panel (b).}
		\label{fig:S05_stars}
	\end{figure}

	After propagation through the two analysis interferometers and before detection, the overall state between signal and idler channels can be described as 
	\begin{equation}
		\begin{aligned}
			\ket{\tilde{\psi}(\varphi)} &= \frac{1}{2\sqrt{2}}[\ket{E_sE_i}+\\
			&+e^{i\varphi_i}\ket{E_sT_i}+e^{i\varphi_s}\ket{T_sE_i}+\\
			&+\left(e^{i(\varphi_s+\varphi_i)}+e^{i\varphi_p}\right)\ket{T_sT_i}+\\
			&+e^{i(\varphi_i+\varphi_p)}\ket{L_sT_i}+e^{i(\varphi_s+\varphi_p)}\ket{T_sL_i}+\\
			&+e^{i(\varphi_s+\varphi_i+\varphi_p)}\ket{L_sL_i}],
		\end{aligned}
		\label{eq}
	\end{equation}
	where subscripts indicating signal and idler channels are explicitly indicated. We would like to stress that states pertaining to second and fourth lines of the equation arise only because photons are probabilistically split at the Franson interferometers inputs and do not carry useful information. This is due to the fact that they can be associated to distinguishable paths being taken by photons. Namely, $\ket{E_sT_i}+\ket{T_sE_i}$ states arise from early photons taking opposite paths on the two twin interferometers, and similarly for the $\ket{L_sT_i}+\ket{T_sL_i}$ states originating from late photons. Recording triple coincidence events (signal, idler and pump) allows to observe all the possible states if displayed in a two-dimensional histogram, which is reported in Fig.~\ref{fig:S05_stars}(a) in a slightly modified version of Fig.~3(a) in the main manuscript. Specifically, diagonal lines are superimposed to the two-dimensional histograms to better illustrate the correspondence between it and its collapsed view in panel Fig.~\ref{fig:S05_stars}(b). By summing counts falling within the ellipses intercepted by each line, from bottom-left to top-right (excluding the black dashed regions), one can build the triple coincidence measurement of Fig.~\ref{fig:S05_stars}(b), which features five peaks \cite{jayakumarTimebinEntangledPhotons2014} and reveals how second and third columns do not carry useful information as they are formed by a superposition of distinguishable, non-interfering states.   
	
\begin{figure}[h!]
		\includegraphics[width=\textwidth]{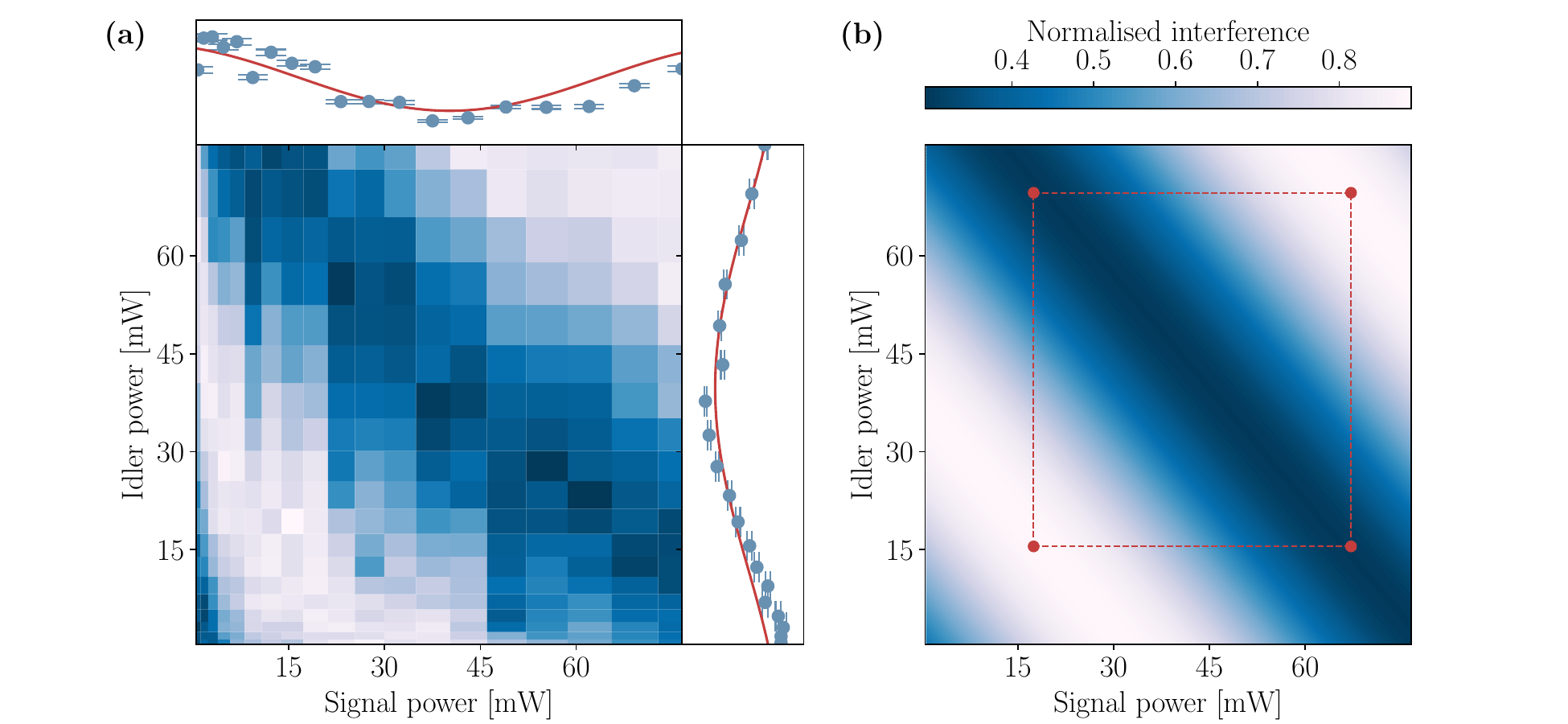}
		\caption{\textbf{Extended Figure 2: Quantum interference calibration map.} (a) Raw data, sliced along horizontal and vertical lines at the centre of the map being displayed in the top/side insets along with slices of the two dimensional fit being performed on the data to obtain the device calibration. (b) Two-dimensional fit, with a dashed rectangle superimposed to the map in order to indicate the path to be followed to span the two qubits Bloch spheres equator.}
		\label{fig:S06_calibration}
	\end{figure}

	Calibration of quantum interference is performed, as described in the main text, by sweeping the electrical power applied to each delay-line phase shifter and monitoring its oscillation as a function of the relative phase. The probability of detecting the state $\ket{T_sT_i}$ can be easily obtained from Eq.~\ref{eq} by projecting the overall state, namely
	\begin{equation}
		P = |\bra{T_sT_i}\tilde{\psi}(\varphi)\rangle|^2 = \frac{1+\cos\left(\varphi_s+\varphi_i-\varphi_p\right)}{4}.
		\label{eq:prob}
	\end{equation} 
	Figure~\ref{fig:S06_calibration}(a) shows raw calibration data as obtained during the measurement, with top and side inset showing horizontal and vertical slices along the centre of the map, respectively. The solid line illustrates the result, on each channel, of a two-dimensional fit to the data based on Eq.~\ref{eq:prob}, with the complete fitted function being displayed in Fig.~\ref{fig:S06_calibration}(b). The dashed rectangular region superimposed to the map illustrates the path to follow in order to span the two qubits Bloch spheres equator. 
	
	\clearpage
	\section*{SM3: Effects impacting visibility}\label{sec:Sup03_visibility}
	Three main effects can be identified to be source of interferometric visibility loss: unbalanced optical losses among the interferometers arms, chromatic dispersion due to the broad photon bandwidth and excessively high photon-pair generation probability per pulse.
	
\begin{figure}[h!]
		\includegraphics[width=0.82\textwidth]{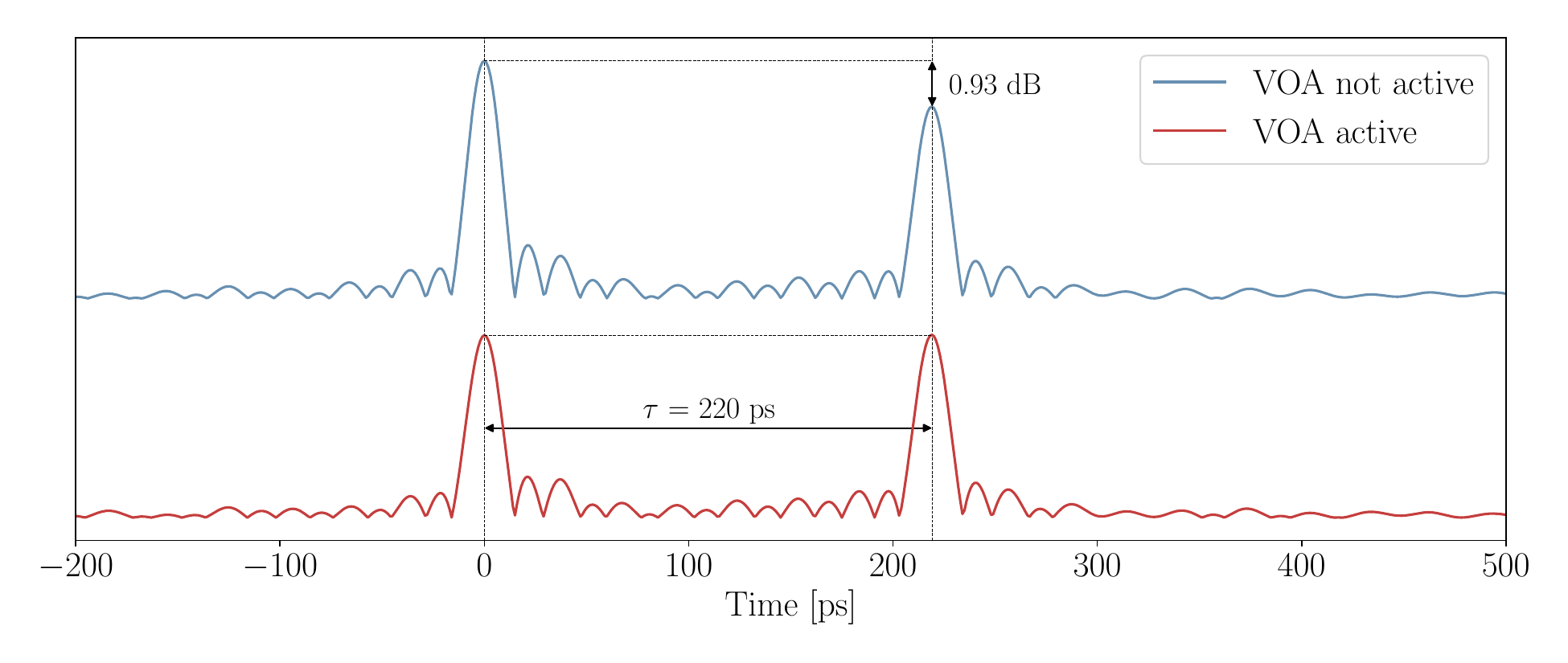}
		\caption{\textbf{Variable optical attenuator operation.} The top trace displays pulse pairs at the interferometer output when the VOA is not biased, while the bottom trace illustrates two balanced pulses due to the tuned splitting ratio.}
		\label{fig:S07_voa}
	\end{figure}

	Optical losses are balanced by using variable optical attenuators, thus by tuning the electrical power, hence the splitting ratio of the Mach-Zehnder interferometers on the short interferometer arm. An example operation of the VOA is displayed in Fig.~\ref{fig:S07_voa}. We couple a femtosecond telecom laser to the devices, pulses are split by interferometer and we use a fast photodiode and oscilloscope to record the intensity traces of the resulting pulses. By tuning the applied electrical power, part of the optical power is damped into the second VOA output port until the two pulses have equal amplitude. From this measurement we estimate an additional propagation loss of approximately \SI{1}{\dB} due to the long spiral waveguide imparting the delay. Ripples around the pulses traces originate from the impulse response of the photodiode.
	
\begin{figure}[h!]
		\includegraphics[width=\textwidth]{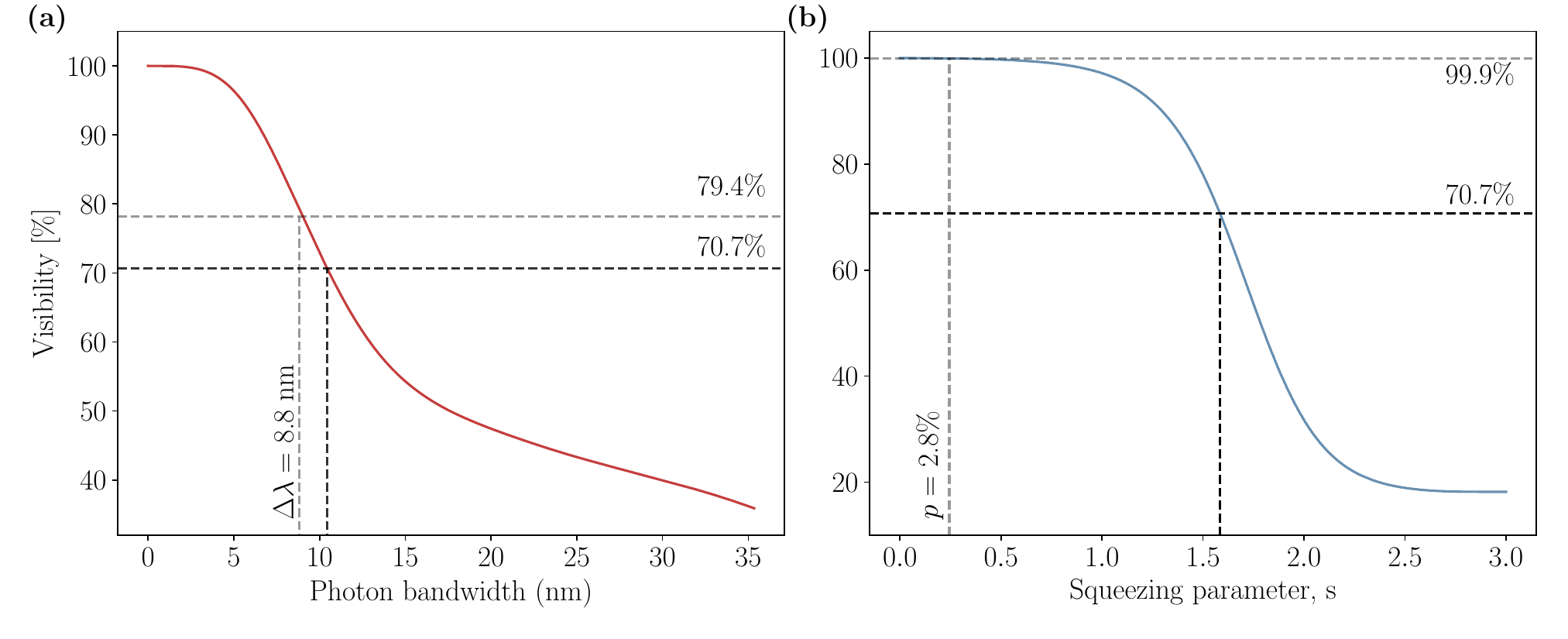}
		\caption{\textbf{Interfering state visibility.} (a) Visibility as a function of signal and idler photon bandwidth. (b) Visibility as a function of the squeezing parameter for the SPDC process. Operating point and threshold for detecting entanglement are marked with dashed lines, confirming that our experiment is conducted at the limit of chromatic dispersion.}
		\label{fig:S08_visibility}
	\end{figure}

	Figure~\ref{fig:S08_visibility} displays the simulated visibility values as a function of photon bandwidth (Fig.~\ref{fig:S08_visibility}(a)) and pumping strength (Fig.~\ref{fig:S08_visibility}(b)). In panel (a) we indicate with dashed lines the visibility threshold for assessing the presence of entanglement, $V=$ \SI{70.7}{\%}, which corresponds to a photon bandwidth of approximately \SI{10.5}{\nm}, and the maximum visibility that we can observe given the full-width at half-maximum of our bandpass filter. This result shows that our measurement approaches the dispersion limit as we observe \SI{78.1}{}$\pm$\SI{2.0}{\%} against the maximum expected of \SI{79.4}{\%}. The simulation is performed by propagating a classical pulse of varying bandwidth across a path extracted from the circuit layout. We account for waveguides bends and anisotropy of the crystal by simulating the optical mode properties as a function of the propagation direction on the plane of our x-cut lithium niobate film. Overlapping of the two dispersed pulses yields then the expected visibility. Deviations from the maximum theoretical visibility are attributed to non-perfectly balanced optical losses in the devices. 
	
	We extract the interferometric visibility as a function of the pumping strength by simulating propagation through the interferometer of states of progressively lower purity. We sweep the probabilities of generating zero- or two-pair states and observe how these affect the measurable visibility by assuming monochromatic fields and lossless propagation. We then model SPDC as a squeezing process of the vacuum field (cfr. \cite{scullyQuantumOptics1997}, Chapters 2 and 16) by applying the squeezing operator 
	\begin{equation}
		\hat{S}(\xi) = \exp\left(\frac{1}{2}\xi^*\cdot\left(\hat{a}\right)^2-\frac{1}{2}\xi^*\cdot\left(\hat{a}^\dagger\right)^2\right)
	\end{equation}
	to the vacuum state $\ket{0}$. The complex squeezing parameter is expressed as $\xi=se^{i\theta}$, with $s\geq0$ being related to the second order nonlinear conversion process via $s\propto\chi^{(2)}A_pL$. $A_p$ is the un-depleted pump field amplitude and $L$ the nonlinear interaction length (cfr. \cite{gerryIntroductoryQuantumOptics2004}, Chapter 9). In this sense, the parameter $s$ is equivalent to the pumping field strength. The resulting squeezed state can be written in the Fock basis as 
	\begin{equation}
		\ket{\xi} = \sqrt{\sech(s)}\sum_{n=0}^{\infty}\frac{\sqrt{(2n)!}}{n!}\left[-\frac{1}{2}e^{i\theta}\tanh(s)\right]^n\ket{2n},
	\end{equation}
	where $n$ indicates the number of photon pairs generated during the down-conversion process (cfr. \cite{gerryIntroductoryQuantumOptics2004}, Chapter 7). The probability of generating $n$ photon pairs as a function of the squeezing parameter is then
	\begin{equation}
		P(s,n) = |\bra{2n}\xi\rangle|^2 = \sech(s)\frac{(2n)!}{2^n(n!)^2}\tanh^{2n}(s).
		\label{eq:prob_s}
	\end{equation} 
	In our experiment, we pump the nonlinear crystal with pulse pairs of known down-conversion probability per pulse, and we can model the entangled state generation by considering the probabilities of generating zero ($p_0$), one ($p_1$) or two ($p_2$) photon pairs per each pair of pulses as
	\begin{equation}
		\begin{aligned}
			& p_0(s) = P(s,0)^2\\
			& p_1(s) = 2 P(s,0)P(s,1)\\
			& p_2(s) = \underbrace{2P(s,0)P(s,2)}_\textrm{two pairs from either pulse} + \underbrace{P(s,1)^2}_\textrm{one pair from each pulse},
		\end{aligned}
	\end{equation} 
	where $P(s,1) = p$ is the measured probability of down-conversion from a pulse, and events with $n>2$ can be neglected at reasonably low pump power. Figure \ref{fig:S08_visibility}(b) reports the calculated visibility, under the above conditions, as a function of the real part of the squeezing parameter, $s$.	We highlight with dashed lines the threshold for CHSH inequality violation and the value of $s$ corresponding to our experimental conditions, thus confirming that the observed visibility is solely limited by chromatic dispersion of broadband photons in the waveguide as effects originating from un-pure state generation can be neglected. Definition of the visibility at $s=0$ has no physical meaning as interference cannot be observed in absence of generated photons.

\end{document}